\documentstyle[12pt]{article}
\newcommand{\beq}{\begin{equation}}
\newcommand{\eeq}{\end{equation}}
\newcommand{\bea}{\begin{eqnarray}}
\newcommand{\eea}{\end{eqnarray}}
\newcommand{\nonu}{\nonumber}
\newcommand{\sma}{\sum_{n=m}^{\infty} \frac{1}{(n-m)!}}
\newcommand{\del}{\delta}
\newcommand{\dl}{\delta^{3}(l'-l)}
\newcommand{\ep}{\epsilon}
\newcommand{\om}{\omega}

\newcommand{\emm}{e_{1}\ldots e_{m}}
\newcommand{\en}{e_{1}\ldots e_{n}}

\newcommand{\dZp}{\frac{\delta^{m}}{\delta u_{m',m}\ldots\delta u_{1',1}} Z_{P'P}}
\newcommand{\dZ}{\frac{\delta^{m}}{\delta u_{m,m}\ldots\delta u_{1,1}} Z_{P'P}}
\newcommand{\nin}{\noindent}

\newcommand{\ama}{A_{PP}(k_{1}',k_{1})}
\newcommand{\amb}{A_{PP}(k_{2}',k_{2})}
\newcommand{\amk}{A_{PP}(k,k)}

\newcommand{\proinm}{\rho_{P}^{in}({\bf k}_{1},\ldots,{\bf k}_{m})}

\newcommand{\proinmbp}{\rho^{in}({\bf k}_{1},\ldots,{\bf k}_{m})}

\newcommand{\proexm}{\rho_{P}^{ex}({\bf k}_{1},\ldots,{\bf k}_{m})}
\newcommand{\proexn}{\rho_{P}^{ex}({\bf k}_{1},\ldots,{\bf k}_{n})}
\newcommand{\proexmbp}{\rho^{ex}({\bf k}_{1},\ldots,{\bf k}_{m})}

\newcommand{\roinm}{d^{in}({\bf k}_{1}',\ldots,{\bf k}_{m}';
{\bf k}_{1},\ldots,{\bf k}_{m})}
\newcommand{\roexm}{d^{ex}({\bf k}_{1}',\ldots,{\bf k}_{m}';
{\bf k}_{1},\ldots,{\bf k}_{m})}
\newcommand{\roexnm}{d^{ex}({\bf k}_{1}',\ldots,{\bf k}_{m}',
{\bf k}_{m+1},\ldots,{\bf k}_{n};{\bf k}_{1},\ldots,{\bf k}_{n})}

\newcommand{\roinmp}{d_{P}^{in}({\bf k}_{1}',\ldots,{\bf k}_{m}';
{\bf k}_{1},\ldots,{\bf k}_{m})}
\newcommand{\roexmp}{d_{P}^{ex}({\bf k}_{1}',\ldots,{\bf k}_{m}';
{\bf k}_{1},\ldots,{\bf k}_{m})}

\newcommand{\pSm}{S_{e_{1}\ldots e_{m}}(k_{1},\ldots,k_{m})}

\newcommand{\pSn}{S_{e_{1}\ldots e_{n}}(k_{1},\ldots,k_{n})}

\newcommand{\zwSSm}{S_{(D')}(k_{1}',\ldots,k_{m}')}
\newcommand{\zwSm}{S_{(D)}(k_{1},\ldots,k_{m})}
\begin{document}
\title{ Multiparton density matrix for the QCD-cascade in DLA approximation
        \thanks{ Work supported in part by the KBN grant: 2 PO 3B 08308
        and by the European Human Capital and Mobility Program ERBCIPDCT940613}
        }
\author{B. Ziaja\\
        Institute of Physics, Jagellonian University\\
        Reymonta 4, 30-059 Cracow, Poland\\
        e-mail: $beataz@ztc386a.if.uj.edu.pl$}
\date{March 1996}
\maketitle
{\abstract The multiparton density matrices of the QCD gluon cascade
are investigated.
The generating functional and master equation for momentum space
multiparton density in the double logarithmic approximation (DLA)
are proposed.
}

\addtolength{\baselineskip}{0.20\baselineskip}

\section{Introduction}

Recently, several groups have analyzed in great detail the multiparton distributions
in the QCD gluon cascades \cite{others}. The results of the investigations show that
perturbative QCD \cite{doksh} provides a powerful framework not only for the
description of
hard quark and gluon jets but also of much softer multiparticle  phenomena.
Although not understood, the hypothesis of parton-hadron duality \cite{lphd} provides
an apparently successful link between theoretical parton distributions and observed
particle spectra. This prescription  was extensively tested in single particle
spectra ( and total multiplicities) and found in a good agreement with available
data (see e.\ g.\ \cite{doksh1}). Recently, there appeared indications that it may also work for
multiparticle correlations \cite{pesch1}. These unquestionable successes invite one
to study further consequences of the theory for processes of particle production.
At this point we would like to notice that, if one wants to exploit fully the
quantum-mechanical aspects of the QCD cascade, it is necessary to study the
multiparton {\bf density matrix}. The multiparticle distributions calculated so far
give only diagonal terms of the density matrix and thus represent rather restricted
( although very important ) part of the information available from the theory.

It is perhaps important to stress that in contrast to what is usually believed,
the interest in studying the multiparticle density matrix is not purely
academic. As suggested in \cite{bk} density matrix allows to obtain the multiparticle
Wigner functions and consequently gives information about the space-time
structure of the system. This allows to make predictions on the
shape and range of the HBT interference with clear experimental consequences.

We present in this paper a study of the
multiparton density matrices of the QCD gluon cascade in the double logarithmic
approximation (DLA) \cite{dla} of QCD. We propose a master equation for the functional
which generates momentum space density matrix of arbitrary order.
The paper is organized as follows. In Section 2 the definition of the density
matrix is given. In Section 3 we recall briefly  the main assumptions of the
DLA approximation for the QCD cascade, and the generating functional method
for multiparticle densities.
In section 4  we propose a scheme to calculate the density matrix
in DLA formalism.
We formulate the master equation for the functional, which can generate
density matrices at first at the most general level, and later in the
approximation of the "quasi - diagonal"
limit. The approximation allows to derive explicit integral equations for
density matrices of arbitrary order. To show explicitly, how the method
works, we derive master equation for single and double inclusive particle
density matrices in the last section.

\section{Definition of the density matrix}

We would like to investigate the properties of the momentum space
density matrix. Let us consider any particle production process where all the
produced particles are already "real" i.e. they are on mass shell.
Then the exclusive m-particle density matrix takes the form~:
\beq
\roexm = \sum_{D',D} \zwSSm \zwSm \prod_{i=1}^{m}(4\om_{ k_{i}'}\om_{k_{i}})^{-1/2},
\eeq
where $ \zwSm $ denotes the probability amplitude to obtain exactly m particles
with the momenta $ k_{1},\ldots, k_{m} $  for the given Feynman diagram D,
and the sum is taken over all possible realization of the process i.e. over all
possible Feynman diagrams for each amplitude separately. We have taken
into account also phase space factors.

The inclusive m-particle density matrix is then given by~:
\bea
\roinm = &         \nonumber\\
\sma \int d_{m+1}[k]_{n} & \roexnm,
\eea
where $d_{i}[k]_{j}\equiv d^{3}k_{i}\ldots d^3k_{j}$.

The "diagonal" elements of the exclusive and inclusive density matrices
give the multiparticle exclusive and inclusive densities respectively~:
\bea
\proexmbp = d^{ex}({\bf k}_{1},\ldots,{\bf k}_{m},;{\bf k}_{1},\ldots,{\bf k}_{m})\\
\proinmbp = d^{in}({\bf k}_{1},\ldots,{\bf k}_{m},;{\bf k}_{1},\ldots,{\bf k}_{m}).
\nonumber
\eea

The on-mass-shell Fourier transforms of the density matrices define
the exclusive and inclusive multiparticle densities in the
configuration space.

\section{Cascade in momentum space. DLA formalism}

In this section we recall the main assumptions of the DLA
approximation in momentum space \cite{dla}.
This approximation accounts for the leading double logarithmic contributions to
multiparton cross section, but it describes quite well the
structure of the parton cascade at high energies. We will shortly review the formalism for
the simplest QCD process: $e^{+}e^{-}\rightarrow q \bar{q}\, + g\ldots g$.
To simplify the analysis one uses here the planar gauge. In this gauge q and
$\bar{q}$ emit soft gluons independently since the interference between two emission
amplitudes vanishes. Virtual corrections appear in the form of the QCD form factors.

\subsection{ Multigluon amplitudes}
For any tree Feynman diagram D ( without 4-gluon vertices) the amplitude to obtain
exactly m gluons (see Fig.\ 1 ) emitted by the initial q ($\bar{q}$) of 4-momentum
$P$ with the account of the virtual corrections reads \cite{doksh}~:
\beq
\pSm=(-1)^{n} e^{-\frac {w(P)}{2}} \prod_{i=1}^{m} M_{P_{i}}(k_{i})\, e^{-\frac {w(k_{i})}{2}},
\eeq
and $M_{P_{i}}(k_{i})$ equals~:
\beq
M_{P_{i}}(k_{i})=g_{S} \frac{(e_{i} \cdot P_{i})}{(k_{i} \cdot P_{i})} \Theta_{P_{i}}(k_{i})\, G_{P_{i}},
\eeq

\begin{tabbing}
where:\\
$g_{S}=\sqrt{4\pi \alpha_{s}}$, \\
n is the number of gluons emitted of quark (antiquark), \\
$k_{i}=(\omega_{i},{\bf k}_{i})$ denotes the 4-momentum of the i-th soft gluon,\\
$e_{i}\equiv e_{i}^{(j)},\,j=0,\ldots,3 $ describes its polarization 4-vector, \\
$P_{i}$ is the 4-momentum of the parent of the ith gluon,\\
$G_{P_{i}}$ denotes the color factor for the given vertex of the tree diagram D,\\
$\Theta_{P_{i}}(k_{i})$ is a generalized step function which restricts the phase
space for $k_{i}$~:\\
\end{tabbing}
\beq
\Theta_{P}(k): \{k^{0} \equiv \omega < P^{0},\, \theta_{{\bf kP}}< \theta, \,
\omega \theta_{{\bf kP}}>Q_{0} \},
\eeq
where $P$ is the momentum of the parent of a given parton $k$, $\theta$ denotes
the emmision angle of the previous parton splitting on the $P-$line, and $Q_{0}$ is a
cut-off parameter.

Radiation factor in (4) contains the function $w(P_{i})$, defined as:~
\beq
w(P)=\int d^{3}k \mid A_{P}(k)\mid ^{2}_{e},
\eeq
where $A_{P}(k)$ introduces the phase space factor $(\omega_{k})^{-1/2}$
in the form~:
\beq
A_{P}(k)=\frac{M_{P}(k)}{\sqrt{2\omega_{k}}},
\eeq
The expression $\mid A_{P}(k)\mid ^{2}_{e}$ in (7) has been  already averaged over
two physical transverse polarizations
$e^{1},e^{2}$. It should be also emphasized that produced gluons are
"real" (on-mass-shell) particles, so the energy $\omega_{k}$ of the gluon of the
momentum k can be approximated as~:
\beq
\omega_{k}=\mid {\bf k}\mid .
\eeq
Summing of the color factors G over the color indices gives the result~:
\beq
G_{P_{i}}G^{*}_{P_{i}}=
\left \{
 \begin{array}{c}
 \mbox{$C_{F}$ dla $P_{i}=P$} \\
 \mbox{$C_{V}$ dla $P_{i}\neq P$,}
 \end{array}
\right.
\eeq
where P denotes the 4-momentum of the quark ( antiquark) which initalizes
the gluon cascade.

\subsection{Multigluon densities}

In the DLA approximation different tree diagrams come from different
non-overlapping kinematic regions. Hence they do not interfere.
Therefore, to calculate exclusive and inclusive multigluon
densities it is enough to sum up the amplitudes (4) incoherently over all possible tree
diagrams D. Hence one gets exclusive density $\proexm$~ in the form~:
\beq
\proexm=\sum_{D} \prod_{i=1}^{m}\mid A_{P_{i}}(k_{i})\mid ^{2}_{e_{i}} \, e^{-w(P_{i})},
\eeq
parametrized by the momentum P of the quark (antiquark) which initializes the
cascade. Multigluon inclusive density $\proinm$ follow from the formula (2), (3)~:
\beq
\proinm=\sma \int d_{m+1}[k]_{n}\,\,\proexn.
\eeq

\subsection{ Generating functional}

Now the difficulty is how to perform the summation in (11) and (12) over all diagrams D
in the convenient way. The problem has been solved by introducing the method
of generating functional (GF) (see \cite{doksh} and references therein).
Generating functional for multigluon densities $Z_{P}[u]$, which fulfills
the master equation~:
\beq
Z_{P}[u]=e^{-w(P)}\, exp(\int d^{3}k \mid A_{P}(k)\mid ^{2}_{e}\, u(k) Z_{k}[u]),
\eeq
reproduces contributions of all possible tree diagrams D and allows to express
multigluon densities $\proinm$ and $\proexm$ as~:
\beq
\proexm=\frac{\delta^{m}}{\delta u_{1}\ldots\delta u_{m}} Z_{P}\mid
_{\{u=0\}}
\eeq
and
\beq
\proinm=\frac{\delta^{m}}{\delta u_{1}\ldots\delta u_{m}} Z_{P}\mid_{\{u=1\}},
\eeq
where $u_{i}$ denotes $u(k_{i})$ and respectively
$ \frac{\del}{\del u_{i}} \equiv \frac{\del}{\del u(k_{i})} $.

We would like to emphasize once more here the simplicity
of description of the multigluon densities (14) and (15) in the language of the GF.
The method allows to forget the complicated summation procedure especially
in the case of inclusive densities, and express the multiplicities
in a simple, compact form (for details see  \cite{wos}).

\section{Calculating of the density matrix in the DLA formalism}

The DLA formalism in momentum space gives a good description of the
structure of the gluon cascade. The GF scheme suggests a clear recipe
how to construct multigluon densities, and allows to describe and investigate
properties of the gluon distributions in a very convenient way.

In this chapter we would like to discuss calculating of the density matrices
in the DLA approximation for the same process namely~:
$e^{+}e^{-}\rightarrow q \bar{q}\, + g\ldots g$.
First we derive the general expression for the exclusive and inclusive density
matrices $\roexmp$ and $\roinmp$.
The task looks more complicated because in this case the interference
between different diagrams in (1) generally
does not vanish. Let us define two GF, $Z_{P}[u]$ and
$Z_{P'}^{*}[w]$, which generate the sum of all the tree amplitudes (4) and
the sum of their complex conjugates respectively~:
\bea
Z_{P}[u]=e^{-w(P)/2}\, exp(\int d^{3}k A_{P}(k) u(k) Z_{k}[u])\nonumber\\
Z_{P'}^{*}[s]=e^{-w(P')/2}\, exp(\int d^{3}k A_{P'}^{*}(k) s(k) Z_{k}^{*}[s]),
\eea
\noindent
The multigluon density matrices can be then expressed as (see Appendix A)~:
\bea
\roexmp = \frac{\delta^{m}}{\delta s_{1'}\ldots\delta s_{m'}}
          \frac{\delta^{m}}{\delta u_{1}\ldots\delta u_{m}}
          Z_{P}[u] Z_{P'}^{*}[s] \mid_{\{u=s=0\}, P=P'}\,\, \nonumber\\
\roinmp = \frac{\delta^{m}}{\delta s_{1'}\ldots\delta s_{m'}}
          \frac{\delta^{m}}{\delta u_{1}\ldots\delta u_{m}}
          Z_{P}[u] Z_{P'}^{*}[s] \mid_{\{u=\frac{\del}{\del s}, s=0\}, P=P'}
\eea

Using the formulae (17) one can obtain complicated equations for the density
matrices. The summation problem does not dissapear. It is only hidden in the
compact form of (17). However, in the above formula we
have taken into account $all$ the possible interferences between different
trees D and D'. Detailed analysis of the DLA  gives the
result that not all the diagrams mix up: one can distinguish classes of
interfering diagrams. Nevertheless, at the general level we did not succeed
in formulation of such a GF which would include only these interfering
diagrams.

However, we do not need the most general formula for
the density matrix. We are interested in its behaviour if the differences
$ \mid k_{1}-k_{1}'\mid ,\ldots, \mid k_{m}-k_{m}'\mid  $ are small ( large momentum differences
will not contribute to Fourier transforms). It can be checked
that in this limit interferences between different diagrams vanish, and one
can sum up only "squared" contributions from the same graphs.
So we construct the new GF for density matrices which reproduces only these
relevant contributions of the tree diagrams and their virtual corrections.
The last step will be to derive explicit integral equations for
single and double particle inclusive density matrix.

\subsection{ Density matrix in quasi-diagonal limit }

We investigate the density matrix in the so-called quasi-diagonal limit
i.e. if for the m-particle matrix the differences
$ \mid k_{1}-k_{1}'\mid
,\ldots, \mid k_{m}-k_{m}'\mid  $ are small.
This requirement simplifies the problem substantively. From the analysis of
all diagrams
contributing to single particle density matrix $d_{P}^{ex}(k_{1}',k_{1})$
it can be proved (see Appendix B) that interference between different diagrams appears only if
$ \mid  \om_{1} - \om_{1'}\mid >>0 $ or $ \mid  \theta_{1P} - \theta_{1'P}\mid >>0 $.
This statement can be generalized for any m-particle density matrix.
If we have m particles, and (m-1) from them are "close" to each other e.g.
$ k_{1}\cong k_{1}',\ldots,k_{m-1}\cong k_{m-1}'$ then the interference
of the different diagrams will take  place only if either energies or
angles of $ k_{m}$ and $k_{m}'$ are strongly ordered.

So, in our approximation we can exclude the interference between the different
diagrams, and sum only the "square" contributions from the same ones. Hence,
the exclusive and inclusive density matrices can be expressed as~:

\bea
\roexmp= \sum_{D} \prod_{i=1}^{m}(4\om_{ k_{i}'}\om_{k_{i}})^{-1/2}\nonumber\\
 \,<S_{\emm}(k_{1}',\ldots,k_{m}')\pSm>_{(\emm)},
\eea
\bea
\roinmp= \sma \sum_{D} \int d_{m+1}[k]_{n} \prod_{i=1}^{m}(4\om_{k_{i}'}\om_{k_{i}})^{-1/2}
\times \nonu\\
\times \prod_{j=m+1}^{n} (4\om_{k_{j}}\om_{k_{j}})^{-1/2}
\,<S_{\en}(k_{1}',\ldots,k_{m}',k_{m+1},\ldots,k_{n})\pSn>_{(\en)}
\eea

The summation in (18), (19) over D can be easily performed using the GF we define
in the next subsection.

\subsection{Generating functional for density matrices in the quasi-diagonal
limit}

We propose a master equation for the new generating functional $Z_{P'P}[u(k',k)]$
in the form~:
\bea
Z_{P'P}[u]=e^{-W(P',P)} \sum_{n=0}^{\infty} \frac{1}{n!}
\int d_{1}[k']_{n}\, d_{1}[k]_{n}
 u(k'_{1},k_{1})\ldots u(k'_{n},k_{n})\times \nonumber\\
\times <A_{P'}^{*}(k'_{1})A_{P}(k_{1})>_{e_{1}}
\ldots <A_{P'}^{*}(k'_{n})A_{P}(k_{n})>_{e_{n}} Z_{k'_{1}k_{1}}[u] \ldots
Z_{k'_{n}k_{n}}[u]\times {\em P}_{1',\ldots,n';1,\ldots,n},
\eea

where the function ${\em P}_{1',\ldots,n';1,\ldots,n}$ introduces the requested parallel angular ordering
for n particles in the form~:

\beq
{\em P}_{1',\ldots,n';1,\ldots,n}=
\sum_{(i_{1},\ldots,i_{n}) \in Perm(1,\ldots,n)}
\Theta( \theta_{k'_{i_{1}}P'}>\ldots>\theta_{k'_{i_{n}}P'})\,\,
\Theta( \theta_{k_{i_{1}}P}>\ldots>\theta_{k_{i_{n}}P}),
\eeq

\noindent
and the radiation factor fulfills the condition~:
\beq
W(P',P)=\frac{w(P')+w(P)}{2}
\eeq
The multiparticle density matrices can be then expressed as (for proof see
Appendix C)~:
\beq
\roexmp =\dZp\mid_{\{u=0\},P=P'}
\eeq
\beq
\roinmp= \dZp\mid_{\{u=\dl\}, P=P'}.
\eeq

If the profile function $u(k',k)$ equals  to $0$ and $\dl$ respectively
the functional $Z_{P'P}$ takes the form~:
\bea
Z_{P'P}[u=0]= & e^{-W(P',P)}\nonumber\\
Z_{PP}[u=\dl]= & 1.
\eea
The GF from (20) is the generalization of the GF defined in (13).
Using GF (20) one can also easily obtain multigluon densities (11)
and (12) in momentum space~:
\bea
\proexm=\dZ\mid_{\{u=0\}, P=P'}\\
\proinm=\dZ\mid_{\{u=\dl\}, P=P'}.\nonumber
\eea

For further analysis we need the exact value of the product of the amplitudes
$<~A_{P'}^{*}(k') A_{P}(k)>_{e}$ which occurs in (20). Averaging over
gluon polarizations (see Appendix D) gives the result~:
\beq
<A_{P'}^{*}(k')A_{P}(k)>_{e} \equiv A_{P'P}(k',k) = 4g_{S}^{2}\, G_{P'}^{*}
G_{P} \frac{1}{\sqrt{4\om^{,3}\om^{3}}}
\frac{1}{\theta_{Pk}\theta_{P'k'}} \Theta_{P'}(k')\Theta_{P}(k)
\eeq

\subsection{ Single and double inclusive density matrix }

From the formula (24) one can derive iterative equations
for inclusive density matrix of an arbitrary order. As an example we present
equations for single and double particle density matrices~:

\bea
d_{P}^{in}(k_{1}';k_{1})=\int d^{3}k A_{PP}(k,k)
d_{k}^{in}(k_{1}';k_{1}) + \nonumber\\
+f_{P}(k_{1}',k_{1}) A_{PP}(k_{1}',k_{1})\,Z_{1',1}[u=\dl]
\eea

\noindent
and
\bea
d_{P}^{in}(k_{1}',k_{2}';k_{1}, k_{2})=
\int d^{3}k \amk d_{k}^{in}(k_{1}',k_{2}';k_{1},k_{2})+
\nonumber\\
+ \ama \frac{\del}{\del u_{2',2}} Z_{1',1}[u=\dl] f_{P}(k_{1}',k_{1})+
\nonumber\\
+ \amb \frac{\del}{\del u_{1',1}} Z_{2',2}[u=\dl] f_{P}(k_{2}',k_{2})+
\nonumber\\
+ \ama Z_{1'1}[u=\dl] \amb Z_{2',2}[u=\dl] \,f_{P}(k_{1}',k_{2}';k_{1},k_{2})+
\nonumber\\
+ \ama Z_{1'1}[u=\dl]\, \int d^{3}k \amk d_{k}^{in} (k_{2}',k_{2}) f_{P}(k_{1}',k;k_{1},k)+
\nonumber\\
+ \amb Z_{2'2}[u=\dl]\, \int d^{3}k \amk d_{k}^{in} (k_{1}',k_{1}) f_{P}(k_{2}',k;k_{2},k)+
\nonumber\\
+ (\int d^{3}k \amk d_{k}^{in}(k_{1}',k_{1}))\,\, (\int d^{3}k \amk d_{k}^{in}(k_{2}',k_{2}))\,\,\,
\eea

\noindent
Factors $f_{P}(k_{1}',\ldots,k_{m}';k_{1},\ldots,k_{m})$ can be expressed as~:
\bea
f_{P}(k_{1}',\ldots,k_{m}';k_{1},\ldots,k_{m})=
e^{-W(P,P)}\sum_{n=0}^{\infty} \frac{1}{n!}
\int d_{1}[p]_{n} A_{PP}(p_{1},p_{1}) \ldots A_{PP}(p_{n},p_{n})\times \nonumber\\
\times {\em P}_{1,\ldots,n,k_{1}',\ldots,k_{m}';1,\ldots,n,k_{1},\ldots,k_{m}}
\eea

For $m=1$ we have calculated $f_{P}(k',k)$ explicitly. It equals to ~:

\beq
f_{P}(k',k)=\Theta(\theta_{k'P}\geq \theta_{kP}) e^{g_{S}^{2}C_{F}
(\ln^{2} \frac{\theta_{kP} P}{ Q_{0}}-\ln^{2} \frac{\theta_{k'P} P}
{ Q_{0}})}  + \Theta(\theta_{k'P} < \theta_{kP}) ( k'\leftrightarrow k )
\eeq


Equations (28), (29) in the limit $k_{1}'=k_{1}, k_{2}'=k_{2}$
reduce to multiparticle density equations \cite{wos} as expected.
However, also for any $k, k'$ the structure of (28), (29) looks  quite
similar to the structure of multiparticle density equations. This suggests
to get the solution using the technique similar to the technique described in
\cite{wos}.

\section{Conclusions}

Our conclusions can be listed as follows~:

1. We have obtained master equation for the density matrices GF. From this
relation one can derive equations for density matrices of arbitrary order.
Integral equations for single and double inclusive densities look quite
similar to the equations for multiparticle densities we already know.
It allows to expect that they can be solved in a way similar to that used
for multiparticle densities.
Hence, as the next step in our approach we will try to derive multiparticle density
explicitly and to investigate their behaviour in the coordinate space.

2. We know that the GF "works" in the quasi-diagonal region but we do not know
what it really means quantitatively. The further quantitative analysis is needed.
If the quasi-diagonal region is large enough, we could easily obtain
the multiparticle densities in the configuration space
taking the on-mass-shell Fourier transform of the density matrix
(it needs also to be checked if the functions we want to transform are
"smooth" enough to get the Fourier integral vanishing in large
$ \mid k_{1}-k_{1}'\mid ,\ldots, \mid k_{m}-k_{m}'\mid  $ limit).

3. One can  try to find the general GF for the density matrices which would
work also far from
the quasi-diagonal region. It would solve the problems mentioned above.
Nevertheless, it seems to be quite complicated. Till now, we did not find
the rule which could iteratively distinguish classes of the interfering diagrams.

\vspace{1.0cm}
\begin{center}
{\bf Acknowledgements}\\
\end{center}
I am very grateful to Professor A. Bialas for many helpful discussions and
suggestions, for critical reading of the manuscript and a continuous interest
throughout this work. I am especially
indebted to Professor J. Wosiek for careful reading the manuscript and
enlightening discussions and comments.

\vspace{1.0cm}

\section{Appendix A}
Functionals  $Z_{P}[u]$ and $Z_{P'}[s]$ defined in (16) generate
the sum of the amplitudes (4) and their complex conjugates over all
possible tree diagrams respectively. It can be seen when one rewrites
i.e. the master equation for $Z_{P}$ in the form of the diagram series
(see Fig.\ 2). From that construction follows the form of the exclusive density matrix
(17)~:
\beq
\roexmp = \frac{\delta^{m}}{\delta s_{1'}\ldots\delta s_{m'}} Z_{P}^{*}[s]\mid_{\{s=0\}}
          \frac{\delta^{m}}{\delta u_{1}\ldots\delta u_{m}} Z_{P}[u]\mid_{\{u=0\}}\,\,
\eeq

Then the inclusive density matrix calculated from (2) and (32) looks like~:
\bea
\roinmp = \sum_{n=0}^{\infty} \frac{1}{n!} \,
( \int d^{3}k \frac{\delta}{\delta s} \frac{\delta}{\delta u} )^{n}
\frac{\delta^{m}}{\delta s_{1'}\ldots\delta s_{m'}} Z_{P}^{*}[s]\mid_{\{s=0\}}\times
\nonu\\
\frac{\delta^{m}}{\delta u_{1}\ldots\delta u_{m}} Z_{P}[u]\mid_{\{u=0\}}\,\,
\eea

\nin and from the identity for the product of any two functionals F, F'~:
\beq
F[u]F'[w]\mid_{\{u=\frac{\del}{\del w}, w=0\}}=
\sum_{n=0}^{\infty} \frac{1}{n!} \,
( \int d^{3}k \frac{\delta}{\delta w} \frac{\delta}{\delta u} )^{n}
F[u]\mid_{\{u=0\}}F'[w]\mid_{\{w=0\}}
\eeq

\nin follows (17).

\section{ Appendix B}

In DLA one gets for the two gluon contributions
four different tree graphs $M_{a}, M_{b}, M_{c}, M_{d}$
defined on non-overlapping kinematic regions (see Fig.\ 3 ).
Emmited gluons are either angular (AO) or energy ordered (EO).

Let us consider all the diagrams contributing to the single particle density matrix
$d_{P}^{ex}(k_{1}',k_{1})$. From the (AO) and (EO) follows that the interference
between any two different graphs will appear only, if either energies $\om_{1},
\om_{1}'$ or emmision angles $\theta_{1P}, \theta_{1'P}$ of produced gluons
are {\sl strongly} ordered (Fig.\ 4).

This statement can be generalized for any m-particle density matrix by induction.
If we have m particles, and (m-1) from them are "close" to each other eg.
$ k_{1}\cong k_{1}',\ldots,k_{m-1}\cong k_{m-1}'$ then the intereference
of the different diagrams will take  place only if either energies or
angles of $ k_{m}$ and $k_{m}'$ are strongly ordered.

\section{Appendix C}

$Z_{P'P}[u(k',k)]$ defined in (20) produces correct exclusive density
matrices (23). This statement follows from the master equation
for $Z_{P'P}$ represented in the form of diagram series (see Fig.\ 5).
The series reproduces all "squared" contributions of the same tree diagrams,
and excludes interference of the different ones
(function ${\em P}_{1',\ldots,n';1,\ldots,n}$).

The form of the inclusive density (24) follows from the formulae (2) and
(23). Substituting (23) into (2) we get as a result~:
\bea
\roinmp = \sum_{n=0}^{\infty} \frac{1}{n!} \,
( \int d^{3}k d^{3}k' \del^{3}(k'-k) \frac{\delta}{\delta u_{k',k}} )^{n}
\times \nonumber\\
\times \frac{\delta^{m}}{ \delta u_{m',m}\ldots\delta u_{1',1}  }
Z_{P'P}[u]\mid_{\{u=0\}}
\eea

which represents the functional $Z_{P'P}[u]$ expanded around "null"
in the "point" $u=\dl$.

\section{Appendix D}

We want to average relation (27) over the physical polarizations of
the produced gluon.
The exact expression to be summed over polarizations looks like~:
\beq
\frac{( e \cdot P) (e' \cdot P')}{(k \cdot P)(k' \cdot P')}
\eeq

\nin
where $e, e'$ are polarization of the same gluon (one gets $e'$ from
$e$ taking the limit $k'=k$). The gauge fixing we use in the approach
allows to neglect contributions to (36) coming from the "nonphysical"
polarizations $e^{0}$ and $e^{3}$.
Furthermore, it requests time components of the physical polarizations
$e^{1}, e^{2}$ to be equal to $0$. Space components of $e^{1}, e^{2}$
can be then constructed in the form~:
\bea
{\bf e}^{2}= \frac{ {\bf P \times k}}{ \mid {\bf P \times k}\mid }\nonu\\
{\bf e}^{1}= \frac{ {\bf e^{2} \times k}}{ \mid {\bf e^{2} \times k}\mid }
\eea

\nin
and respectively for $e'$~:
\bea
{\bf e}^{,2}= \frac{ {\bf P' \times k'}}{ \mid {\bf P' \times k'}\mid }\nonu\\
{\bf e}^{,1}= \frac{ {\bf e^{,2} \times k'}}{ \mid {\bf e^{,2}\times k'}\mid }
\eea

Summing over these 2 polarizations and expanding scalar products in (36)
one obtains finally~:
\beq
\sum_{j=1,2}\frac{( e^{j} \cdot P) (e^{,j} \cdot P')}{(k \cdot P)(k' \cdot P')}=
\frac{4}{\om \om'} \,\, \frac{1}{\theta_{Pk} \theta_{P'k'}}
\eeq

The result can be easily confirmed for any two physical polarizations
${\bf \ep}^{1}, {\bf \ep}^{2}$ lying in the plane ${\bf e}^{1} {\bf e}^{2}$~:
\bea
{\bf \ep}^{1}=  cos \varphi \, {\bf e}^{1} + sin \varphi \, {\bf e}^{2}\nonu\\
{\bf \ep}^{2}= -sin \varphi \, {\bf e}^{1} + cos \varphi \, {\bf e}^{2}
\eea

\nin and respectively for ${\bf \ep}^{,1}, {\bf \ep}^{,2}$ with the phases
$\varphi=\varphi '$.

\end{document}